\documentclass[showpacs,aps,pre,twocolumn,superscriptaddress,textsuperscript]{revtex4}
\usepackage{graphicx}
\usepackage{amssymb}
\usepackage[sort&compress]{natbib}
\begin{document}
\title {Graphene-like quaternary compound SiBCN: a new wide direct band gap semiconductor predicted by a first-principles study}
\author{Yan Qian}
\email[Corresponding author:]{qianyan@njust.edu.cn}
\affiliation{Department of Applied Physics, Nanjing University of Science and Technology, Nanjing 210094, China}
\author{Haiping Wu}
\email[Corresponding author:]{mrhpwu@njust.edu.cn}
\affiliation{Department of Applied Physics, Nanjing University of Science and Technology, Nanjing 210094, China}
\author{Erjun Kan}
\affiliation{Department of Applied Physics, Nanjing University of Science and Technology, Nanjing 210094, China}
\author{Kaiming Deng}
\affiliation{Department of Applied Physics, Nanjing University of Science and Technology, Nanjing 210094, China}

\date{\today}
\begin{abstract}
Due to the lack of two-dimensional silicon-based semiconductors and the fact that most of the components and devices are generated on single-crystal silicon or silicon-based substrates in modern industry, designing two-dimensional silicon-based semiconductors is highly desired. With the combination of a swarm structure search method and density functional theory in this work, a quaternary compound SiBCN with graphene-like structure is found and displays a wide direct band gap as expected. The band gap is of $\sim$2.63 eV which is just between $\sim$2.20 and $\sim$3.39 eV of the highlighted semiconductors SiC and GaN. Notably, the further calculation reveals that SiBCN possesses high carrier mobility with $\sim$5.14$\times$10$^{3}$ and $\sim$13.07$\times$10$^{3}$ cm$^{2}$V$^{-1}$s$^{-1}$ for electron and hole, respectively. Furthermore, the ab initio molecular dynamics simulations also show that the graphene-like structure of SiBCN can be well kept even at an extremely high temperature of 2000 \textit{K}. The present work tells that designing multicomponent silicides may be a practicable way to search for new silicon-based low-dimensional semiconductors which can match well with the previous Si-based substrates.
\end{abstract}
\maketitle
\section{Introduction}
 With the development of modern industry, electronic devices need more and more smaller. Low-dimensional materials are just the ones that can satisfy this demand\cite{Boon,Tang,Rao}. Besides, the properties of low-dimensional materials are always different from those of the bulk counterparts. For instance, graphite behaves as common metallic nature, while graphene shows a zero-gap semiconductor with Dirac points\cite{Novoselov1,Novoselov2}. Subsequently, motivated by the discovery of graphene, many researchers focus on exploring new low-dimensional monolayered materials via theoretical or experimental methods. As a result, many such materials have been reported, for example, h-BN sheet, borophene, phosphorene, germanene, silicene, 2D transition metal dichalcogenides, and 2D transition metal dinitride, etc\cite{Pacil, Andrew, Cahangirov, Ding, Liu, Xu, Chen, Fang, Wu}.

On the other hand, semiconductors are so important for modern industry, such as integrated circuit, power component, light-emitting diode (LED) fields and so on, thus searching for two-dimensional (2D) semiconducting materials shows very necessary and attracts a lot of research interests. Hexagonal boron nitride (h-BN) with a gap of 5.8 eV is a representative one\cite{Rubio}. Atomically thin MoS$_{2}$ was also predicted as a direct-gap semiconductor in 2010\cite{Kin}. Few-layer black phosphorus was theoretically reported as a semiconductor with a direct-gap which could increase up to 2.0 eV by decreasing the number of layers to one.\cite{Tran} Recently in 2016, atomically thin InSe was reported as an indirect band gap semiconductor with a gap of $\sim$1.50 eV\cite{Denis}. However, despite many 2D semiconductors were theoretically or experimentally found continuously, few literatures have reported on applications of these 2D semiconductors in the above mentioned fields. We think the main reason is that semiconducting materials are usually grown on Si or silicide substrates to date, while the most reported 2D semiconductors can not match well with these substrates. Most recently, SiC and GaN become the highlighted semiconducting materials attributed to their wide band gaps. However, SiC is an indirect bang gap semiconductor, single-crystal GaN is so expensive and there is still no suitable substrate for its growth.

 The above fact encourages researchers to search for 2D silicon-based semiconductors. Silicene, as a 2D material with a buckling hexagonal honeycomb structure, was theoretically designed in 2009\cite{Cahangirov}. The electronic calculation shows it exhibits gapless semimetal, instead of semiconducting like its bulk counterpart, and its charge carriers behave like a massless Dirac fermion due to linear band crossing at the Fermi energy level (\textit{E}$_{F}$). Following this prediction, some work was focused on tuning the band gap of silicene. For example, Du \textit{et al} regulated silicene from semimetal to semiconductor by chemically adsorbing oxygen atoms, and the band gap can be ranged from 0.11 to 0.30 eV by tuning the adsorption configurations and amount of adsorbed oxygen atoms\cite{Yi}. But whether silicene with the oxygen atoms adsorbed on surface could match well with other silicon-based devices is an open question, because the adsorbed atoms would destroy the structure at the atomic level. Additionally, chemical modification is still hard to be achieved by experiment now. Besides, Ni \textit{et al} also theoretically predicted that silicene could be tuned to a semiconductor via a vertical electric field\cite{Ni}. However, the electric field applied to open an effective gap is too high to achieve in experiment, especially in practical applications, for example, an electric field of 0.16 V/{\AA} is required to open a gap of 0.026 eV.

Therefor, searching for intrinsic semiconducting 2D silicides, instead of tuning silicene by different means, perhaps can pave a way. Originated from this idea, we theoretically designed a new wide direct band gap semiconductor - a hexagonal single-atom-thick quaternary compound SiBCN in this work.

\section{Computational methods}
The structures employed herein are predicted by using the swarm-intelligence based structural prediction calculations implemented in CALYPSO code. The code is developed to search for the stable structures of compounds\cite{wang1,wang2}. The underlying ab initio structural relaxations and electronic band structure calculations are carried out in the framework of density functional theory (DFT) within generalized$-$gradient approximations using the Perdew$-$Burke$-$Ernzerhof (PBE) exchange$-$correlation functional and projector augmented wave (PAW) potentials\cite{Perdew, Kresse1} as played in VASP code\cite{Kresse2}. The structural relaxations are performed until the Hellmann-Feynman force on each atom reduces by less than 0.001 eV/{\AA}. To ensure high accuracy, the k-point density and the plane waves cutoff energy are increased until the change of the total energy is less than 10$^{-5}$ eV, and finally the Brillouin$-$zone (BZ) integration is carried out using 15$\times$15$\times$1 Monkhorst$-$Pack grid in the first BZ, the plane waves with the kinetic energy up to 600 eV is employed. In addition, the simulations are performed using a 2$\times$2$\times$1 supercell based on unit cell, and the repeated layered geometry is with a thick vacuum region of 20 {\AA}.
\section{Results and discussions}
The four most energetically favorable ones among the predicted structures are discussed here. Their total energies are listed in TABLE I, and the corresponding optimized geometries are plotted in Fig. 1. The table tells that Struc1 has the lowest total energy with the value at least 2.18 eV per primitive cell lower than those of the other structures, indicating Struc1 is the ground-state structure for SiBCN. The result can be clearly explained by the arrangement of the composed atoms. From Fig. 1, it can be concluded that no Si-B or C-N bonds are formed in Struc1, which is specially different with the other structures. According to the electronegativity of the four elements, the formation energies of Si-B and C-N bonds are higher than the other types of bonds, leading to Struc1 possessing the lowest total energy. In detail, it clearly shows from Fig. 1(a) that every atom in Struc1 is bonded by other two atoms and a graphene-like single-atom-thick structure is formed. The lengthes of Si-C, Si-N, B-C, B-N bonds are 1.75, 1.73, 1.52, 1.50 {\AA}, respectively. In order to compare with this configuration, the structural properties of the corresponding binary compounds SiC, Si$_{3}$N$_{4}$, B$_{4}$C, and h-BN are calculated as well, and the calculated bond lengthes agree well with the experimental ones, as listed in TABLE II. It indicates from the table that Si-C bond length in Struc1 is shorter than 1.89 {\AA} in bulk SiC\cite{Gomes}, and Si-N bond length is consistent with 1.72 {\AA} in bulk Si$_{3}$N$_{4}$\cite{Hardie}, while the lengthes of B-C and B-N bonds are much larger than 1.43 and 1.44 {\AA} in B$_{4}$C\cite{Allen} and h-BN\cite{Jin}, respectively. The shortening of Si-C bond in SiBCN comparing with that in bulk SiC can be explained by the fact that radius of B and N atoms are much shorter than that of Si, thus replacing Si with B and N atoms would induce the compression of SiC volume, resulting in the shortening of Si-C bond. For the lengthes of B-C and B-N bonds, the situation is just on the contrary. The lengthes of Si-C, Si-N bonds decrease from 1.75 to 1.73 {\AA}, caused by the shortening of atomic radius from C to N. This rule can also explain the fact that the length of B-N bond is shorter than that of B-C bond. The difference of bond lengthes drives slight deformation of hexagonal structure. From the side-view, Struc1 presents an absolutely flat structure, namely no buckling is built. This attributes to no formation of Si-Si bond.

\begin{table}[!hbp]
\caption{The total energies (in eV per primitive cell) of the four most energetically favorable structures relative to that of Struc1, and the total energy of Struc1 has been shifted to zero.}
\begin{tabular}{p{1.6cm}p{1.6cm}p{1.6cm}p{1.6cm}}
\hline
\hline
Struc1 &Struc2 & Struc3 & Struc4 \\
\hline
0 & 2.18 & 3.24 & 3.38  \\
\hline
\hline
\end{tabular}
\end{table}

\begin{figure}[htbp]
\centering
\includegraphics[width=8.5cm]{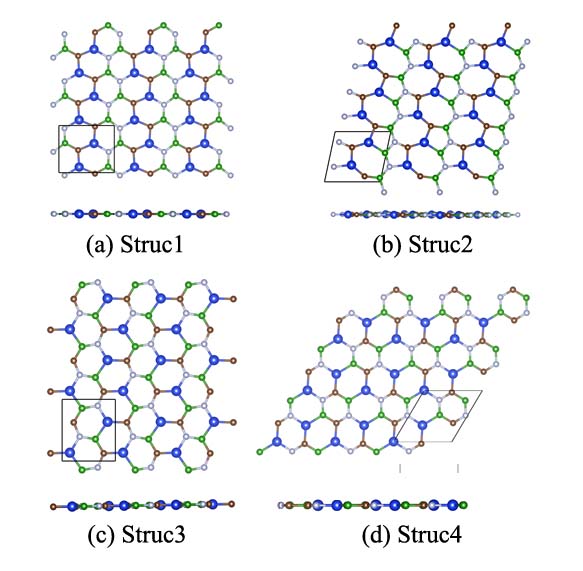}
\caption{(Color online). The top and side views of the four energetically favorable structures of SiBCN. The blue, green, brown, and gray spheres present Si, B, C, and N atoms, respectively.}
\label{fig:Figure1}
\end{figure}

\begin{table}[!hbp]
\caption{The bond lengthes (in \AA) in Struc1 SiBCN and the corresponding binary compounds given by experiment and our calculations.)}
\begin{tabular}{p{1.7cm}p{1.6cm}p{1.6cm}p{1.6cm}p{1.6cm}}
\hline
\hline
      &Si-C & Si-N & B-C & B-N\\
\hline
SiBCN & 1.75 & 1.73 & 1.52 & 1.50  \\
\hline
SiC (exp$\verb|/|$cal) & 1.89\cite{Gomes} $\verb|/|$1.89 & - & - & - \\
\hline
Si$_{3}$N$_{4}$ (exp$\verb|/|$cal) & - & 1.72\cite{Hardie} $\verb|/|$1.73 & - & - \\
\hline
B$_{4}$C (exp$\verb|/|$cal) & - & - & 1.43\cite{Allen} $\verb|/|$1.44  & - \\
\hline
h-BN (exp$\verb|/|$cal) & - & - & - & 1.44\cite{Jin} $\verb|/|$1.44 \\
\hline
\hline
\end{tabular}
\end{table}

Owing to the result that Struc1 is the ground-state structure of SiBCN, the thermodynamic stability of this SiBCN structure is discussed through the cohesive energy. The cohesive energy is
defined as the follows:
\begin{equation}
\Delta E=(2E_{Si}+2E_{B}+2E_{C}+2E_{N}-E_{SiBCN})/8
\label{eq:fn}
\end{equation}
where \textit{E}$_{SiBCN}$ is the total energy of one SiBCN primitive cell, \textit{E}$_{Si}$, \textit{E}$_{B}$, \textit{E}$_{C}$ and \textit{E}$_{N}$ are the total energies of the isolated single Si, B, C and N atoms, respectively. The calculated cohesive energies, including those of graphene, h-BN, and silicene for comparison, are listed in Table III. The cohesive energy of SiBCN is 6.35 eV per atom. This value is a bit lower than those of graphene and h-BN, but is much higher than that of silicene. This is consistent with the fact that graphene and h-BN can be freestanding, while silicene must be grown on some matched substrates. The relatively high cohesive energy of SiBCN demonstrates that the graphene-like structure could exist stably.

\begin{table}[!hbp]
\caption{The cohesive energies (in eV per atom) of Struc1 SiBCN and three representative 2D materials by our calculations.}
\begin{tabular}{p{1.6cm}p{1.6cm}p{1.6cm}p{1.6cm}}
\hline
\hline
SiBCN & Graphene & h-BN &  Silicene  \\
\hline
6.35 & 8.01 & 7.08 &  3.98  \\
\hline
\hline
\end{tabular}
\end{table}

To further evaluate the thermal stability of SiBCN, the ab initio molecular dynamics (AIMD) simulations were performed. During the calculations, a large 4$\times$4$\times$1 supercell based on primitive cell is employed, AIMD simulations are calculated using the NVT ensembles, the temperature controlled by the Nos\'{e}-Hoover method is ranged from 300 K to 2000 K, and the simulations are lasted for 10 ps with a time step of 2.0 fs. Simulation snapshots of the last step for SiBCN at different temperature are described in Fig. 2. From the top and side views, it clearly shows that there is no breaking of the bonds and the original configuration is well kept even at the high temperature of 2000 K. The above performance means that single-atom-layer SiBCN possesses high thermal stability, indicating their potential applications even at extremely high temperature.

\begin{figure}[htbp]
\centering
\includegraphics[width=8.5cm]{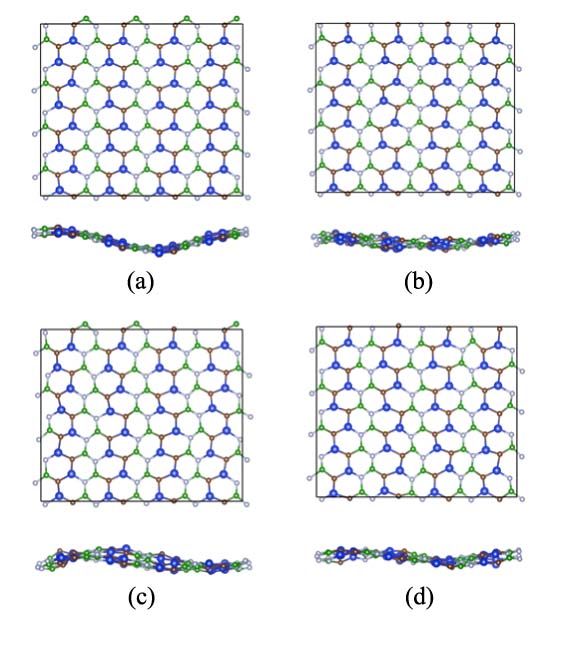}
\caption{(Color online). Snapshots of Struc1 SiBCN equilibrium structures at (a) 300 K, (b) 1000 K, (c) 1500 K, and (d) 2000 K at the last step of 10 ps AIMD simulations.}
\label{fig:Figure2}
\end{figure}

Encouraged by the stability of SiBCN, the electronic property of this new single-atom-thick material is uncovered furthermore. The density of electronic states (DOS) and partial density of electronic states (PDOS) are described in Figs. 3(a) and 3(b). The DOS clearly shows that the SiBCN behaves as semiconductor with a wide energy gap of $\sim$2.63 eV which is just between 2.20 and 3.39 eV for SiC and GaN, respectively\cite{Morko}. This indicates SiBCN can be a potential candidate to be utilized in the corresponding fields SiC and GaN applied. The PDOS presents that the valence band is mainly composed of C 2\textit{p} states and some of N 2\textit{p} states, while the conduction band is mainly composed of B 2\textit{p} and Si 3\textit{p} states and some of N 2\textit{p} states. In the energy range below \textit{E}$_{F}$, there are many PDOS peaks of the four elements located at the same energy levels. This performance demonstrates that there are strong interaction among the neighboring atoms, which is confirmed by the electron localization functions (ELF) described in Fig. 4. The figure clearly presents that the strong bonds are formed between the nearest neighboring atoms, and all the bonds have $\sigma$ characteristic. The formation of strong bonds can well explain why the compound still remains the original structure even at extremely high temperature.

\begin{figure}[htbp]
\centering
\includegraphics[width=8.5cm]{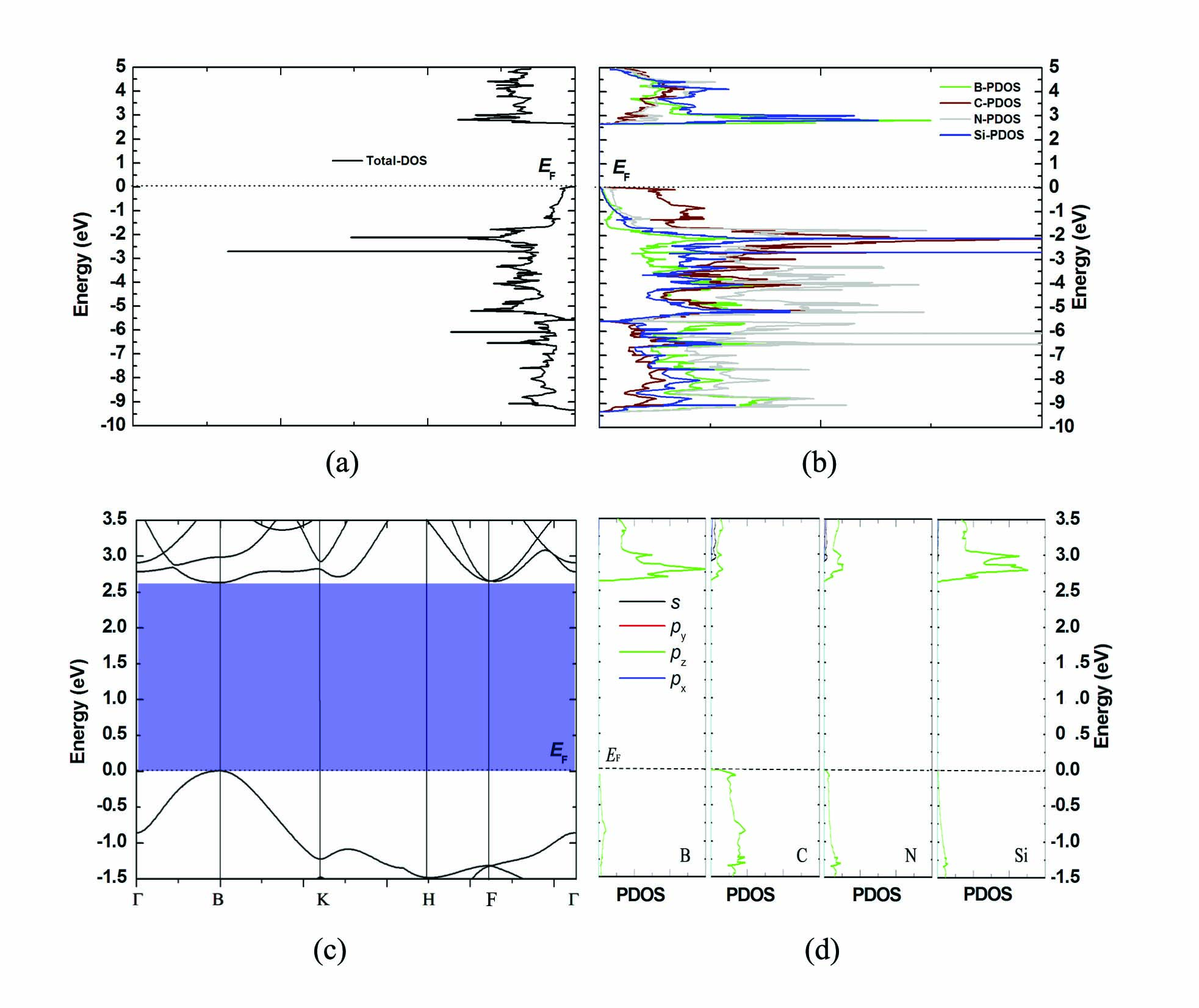}
\caption{(Color online). (a) The density of electronic states (DOS) and (b) partial density of electronic states (PDOS) of Struc1 SiBCN. (c) The band structure of Struc1 SiBCN. (d) The projected density of electronic states B, C, N, and Si, respectively.}
\label{fig:Figure3}
\end{figure}
\begin{figure}[htbp]
\centering
\includegraphics[width=8.5cm]{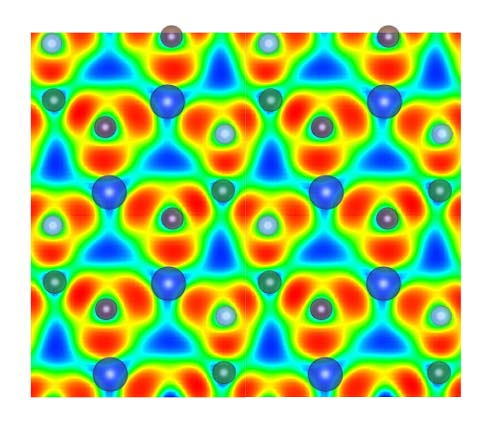}
\caption{(Color online). The electron localization functions of Struc1 SiBCN.}
\label{fig:Figure4}
\end{figure}

Since semiconductors with a direct band gap are more suitable for the applications in photovoltaic components, the band structure of SiBCN is explored next. The band dispersion curves are plotted in Fig. 3(c). From the figure, it can be clearly concluded that the valence band maximum (VBM) and the conduction band minimum (CBM) are situated at the same reciprocal point \textit{B}. This elucidates that SiBCN is a direct band gap semiconductor. Additionally, Fig. 3(d) also shows that the valence band is mainly composed of C and N 2\textit{p$_{z}$} state and the conduction band is mainly composed of B 2\textit{p$_{z}$} and Si 3\textit{p$_{z}$} states, respectively. This characteristic can be confirmed by the partial (band decomposed) charge density of valence and conduction band edges, as plotted in Fig. 5. Fig. 5(a) shows that the charge is mainly located on C atoms in valence band, and the charge is mainly situated on Si and B atoms in conduction band.

\begin{figure}[htbp]
\centering
\includegraphics[width=8.5cm]{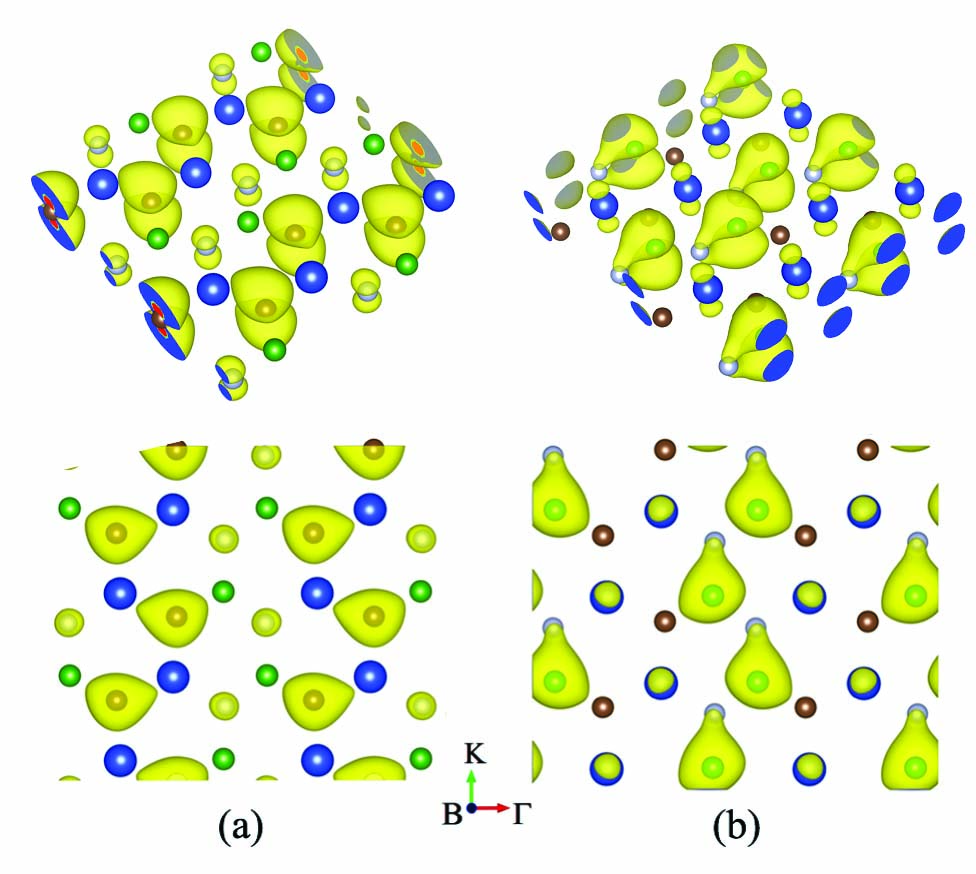}
\caption{(Color online). The partial (band decomposed) charge density of (a) valence and (b) conduction band edges of Struc1 SiBCN.}
\label{fig:Figure5}
\end{figure}

Finally, the carrier effective mass is evaluated due to its importance for the practical application of semiconductors. Here the electron effective mass (m$_{e}$$^{*}$) at CBM and hole effective mass (m$_{h}$$^{*}$) at VBM are defined by the following expression
\begin{equation}
m^{*}=\hbar^{2}(\partial^{2}E/\partial k^{2})^{-1}
\label{eq:fn}
\end{equation}
where \textit{E} and \textit{k} are the energy and reciprocal lattice vector along monolayer. The calculated m$_{e}$$^{*}$ is 3.79 and 0.45 m$_{0}$ along the B-$\Gamma$ and B-K directions, respectively, and the m$_{h}$$^{*}$ is 1.48 and 0.33 m$_{0}$ along the two directions (m$_{0}$ denotes the free-electron mass.). It is clearly that both m$_{e}$$^{*}$ and m$_{h}$$^{*}$ along B-$\Gamma$ direction is much heavier than the ones along B-K direction. This carrier effective mass anisotropy can be explained by partial charge density of valence and conduction band edges in Fig. 5. The figure indicates that the charge prefers expanding along B-K direction to along B-$\Gamma$ direction, telling that the carrier transporting along B-K direction is more easily than along B-$\Gamma$ direction. A small effective mass usually represents high carrier mobility, this encourages us to explore the charge carrier transport property furthermore. The phonon limited carrier mobility in 2D materials can be calculated by the definition\cite{Bruzzone,Takagi,Fiori,Qiao,Jia}

\begin{equation}
\mu_{2D}=\frac{e\hbar^{3}C_{2D}}{k_{B}Tm^{*}m_{a}(E_{l}^{i})^{2}}
\label{eq:fn}
\end{equation}

where e is the electron charge and $\hbar$ is Planck's constant divided by 2$\pi$, \textit{k}$_{B}$ is Boltzmann's constant, and \textit{T} is the temperature and 300 \textit{K} is employed here. m$^{*}$ is the effective mass in the transport direction, and m$_{a}$ is the average effective mass determined by m$_{a}$ = (m$^{*}$$_{B-K}$m$^{*}$$_{B-\Gamma}$)$^{1/2}$, \textit{C}$_{2D}$ is the elastic modulus of the longitudinal strain in the propagation directions of the longitudinal acoustic wave and is defined by (\textit{E}-\textit{E}$_{0}$)/\textit{S}$_{0}$=\textit{C}$_{2D}$($\Delta$\textit{l}/\textit{l}$_{0}$)$^{2}$/2, here \textit{E} is the total energy and \textit{S}$_{0}$ is the lattice volume at equilibrium for a 2D system. \textit{E}$_{l}$$^{i}$ represents the deformation potential constant of VBM for hole or CBM for electron along the transport direction and is expressed by \textit{E}$_{l}$$^{i}$=$\Delta$\textit{V}$_{i}$/($\Delta$\textit{l}/\textit{l}$_{0}$), where $\Delta$\textit{V}$_{i}$ is the energy change of the \textit{i}$^{th}$ band under proper cell compression and dilatation, \textit{l}$_{0}$ and $\Delta$\textit{l} are the lattice constant in the transport direction and the deformation of \textit{l}$_{0}$, respectively. According to this equation, the evaluated electron mobility is $\sim$5.14$\times$10$^{3}$cm$^{2}$V$^{-1}$s$^{-1}$ along B-K direction and $\sim$0.61$\times$10$^{3}$cm$^{2}$V$^{-1}$s$^{-1}$ along B-$\Gamma$ direction, the hole mobility is calculated as $\sim$13.07$\times$10$^{3}$cm$^{2}$V$^{-1}$s$^{-1}$ along B-K direction and $\sim$2.91$\times$10$^{3}$cm$^{2}$V$^{-1}$s$^{-1}$ along B-$\Gamma$ direction, respectively. This mobility is much higher than $\sim$58.80cm$^{2}$V$^{-1}$s$^{-1}$ of Boron Nitride Nanoribbons\cite{HZ} and $\sim$0.3 cm$^{2}$V$^{-1}$s$^{-1}$ of MoS$_{2}$\cite{KS}, and is of the same order of magnitude as $\sim$$\times$10$^{3}$cm$^{2}$V$^{-1}$s$^{-1}$ of atomically thin InSe\cite{Denis}. High carrier mobility indicates that SiBCN could be a suitable material for high efficiency solar cell.

\section{Conclusions}
In conclusion, a new wide direct band gap semiconductor SiBCN is predicted by using first principles calculations in combination with a swarm structure search method. The results show that SiBCN has a graphene-like structure and every atom is bonded with the other types of atoms, and this atomic configuration can be kept even at extremely high temperature. The further investigation also tells that SiBCN exhibits high carrier mobility. Due to possessing a wide direct band gap and high carrier mobility, SiBCN can be utilized in modern industry, such as integrated circuit, power component, blue LEDs, solar cell, etc.

\vspace{1ex}
\begin{acknowledgments}
This work was supported by the National Natural Science Foundation of China (Grant nos. 11404168, 11304155, and 11374160).
\end{acknowledgments}


\begin{thebibliography}{99}
\bibitem{Boon}K. T. Boon, and X. H. Sun, Chem. Rev. {\bf 107}, 1454 (2007).
\bibitem{Tang}Q. Tang, and Z. Zhou, Progress in Materials Science {\bf 58}, 1244 (2013).
\bibitem{Rao} C. N. R. Rao, K. Gopalakrishnan, and U. Maitra, ACS Appl. Mater. Interfaces {\bf 7}, 7809 (2015).
\bibitem{Novoselov1} K. S. Novoselov, A. K. Geim, S. V. Morozov, D. Jiang, Y. Zhang, S. V. Dubonos, I. V. Grigorieva, and A. A. Firsov, Science {\bf 306}, 666 (2004).
\bibitem{Novoselov2} K. S. Novoselov, A. K. Geim, S. V. Morozov, D. Jiang, M. I. Katsnelson, I. V. Grigorieva, S. V. Dubonos, and A. A. Firsov, Nature {\bf 438}, 197 (2005).
\bibitem{Pacil} D. Pacil\'{e}, J. C. Meyer, \c{C}. \"{O}. Girit, and A. Zettl, Appl. Phys. Lett. {\bf 92}, 133107 (2008).
\bibitem{Andrew} J. M. Andrew , X. F. Zhou, B. Kiraly, J. D. Wood, D. Alducin, B. D. Myers, X. L. Liu, B. L. Fisher, U. Santiago, J. R. Guest, M. J. Yacaman, A. Ponce, A. R. Oganov, M. C. Hersam, and N. P. Guisinger, Science {\bf 6267}, 1513 (2015).
\bibitem{Cahangirov} S. Cahangirov, M. Topsakal, E. Akt\"{u}rk, H. Sahin, and S. Ciraci, Phys. Rev. Lett. {\bf 102}, 236804 (2009).
\bibitem{Ding} Y. Ding and J. Ni, Appl. Phys. Lett. {\bf 95}, 083115 {2009}.
\bibitem{Liu} H. Liu, A. T. Neal, Z. Zhu, Z. Luo, X. F. Xu, D. Tom\'{a}nek, and P. D.Ye, ACS Nano {\bf 8}, 4033 (2014).
\bibitem{Xu} M. S. Xu, T. Liang, M. M. Shi, and H. Z. Chen, Chem. Rev. {\bf 113}, 3766 (2013).
\bibitem{Chen} L. Chen, C. C. Liu, B. J. Feng, X. Y. He, P. Cheng, Z. J. Ding, S. Meng, Y. G. Yao, and K. H. Wu, Phys. Rev. Lett. {\bf 109}, 056804 (2012).
\bibitem{Wu} H. P. Wu, Y. Qian, R. F. Lu, and W. S. Tan, Phys. Lett. A {\bf 380}, 768 (2016).
\bibitem{Fang} F. Wu, C. X. Huang, H. P. Wu, C. Lee, K. M. Deng, E. J. Kan, and P. Jena, Nano Lett. {\bf 15}, 8277 (2015).
\bibitem{Rubio} A. Rubio, J. L. Corkill, andM. L. Cohen, Phys. Rev. B {\bf 49}, 5081 (1994).
\bibitem{Kin} K.F. Mak, C.G. Lee, J. Hone, J. Shan, and T.F. Heinz, Phys. Rev. Lett. {\bf 105}, 136805 (2010).
\bibitem{Tran} Vy Tran, Ryan Soklaski, Yufeng Liang, and Li Yang, Phys. Rev. B {\bf 89}, 235319 (2014).
\bibitem{Denis} Denis A. Bandurin, Anastasia V. Tyurnina, Geliang L. Yu, Artem Mishchenko, Viktor Z\'{��}lyomi, Sergey V. Morozov, Roshan Krishna Kumar, Roman V. Gorbachev, Zakhar R. Kudrynskyi, Sergio Pezzini, Zakhar D. Kovalyuk, Uli Zeitler, Konstantin S. Novoselov, Amalia Patan\`{e}, Laurence Eaves, Irina V. Grigorieva, Vladimir I. Fal'ko, Andre K. Geim and Yang Cao, Nat. Nanotech. doi:10.1038/nnano.2016.242
\bibitem{Yi} Y. Du, I.C. Zhuang, H.S. Liu, X. Xu, S. Eilers, K.H. Wu, P. Cheng, J.J. Zhao, X.D. Pi, K.W. See, G. Peleckis, X.L. Wang, and S.X. Dou, ACS Nano, {\bf 8}, 10019 (2014).
\bibitem{Ni}Z.Y. Ni, Q.H. Liu, K.C. Tang, J.X. Zheng, J. Zhou, R. Qin, Z.X. Gao, D.P. Yu, and J. Lu, Nano Lett. {\bf 12}, 113 (2012).
\bibitem{wang1} Y. Wang, J. Lv, L. Zhu, and Y.M. Ma, Phys. Rev. B {\bf 82}, 094116 (2010).
\bibitem{wang2} Y. Wang, J. Lv, L. Zhu, and Y.M. Ma, Comput. Phys. Commun. {\bf 183}, 2063 (2012).
\bibitem{Perdew} J.P. Perdew, K. Burke, and M. Ernzerhof, Phys. Rev. Lett. {\bf 77}, 3865 (1996).
\bibitem{Kresse1} G. Kresse and D. Joubert, Phys. Rev. B: Condens. Matter Mater. Phys. {\bf 59}, 1758 (1999).
\bibitem{Kresse2} G. Kresse, and J. Furthm\"{u}ller, Comput. Mater. Sci. {\bf 6}, 15 (1996).
\bibitem{Gomes} A.H. MESQUITA, Acta Cryst. {\bf 23}, 610 (1967).
\bibitem{Hardie} D. Hardie and K. H. Jack, Nature {\bf 180}, 332, (1957).
\bibitem{Allen} A.C. Larson, AIP Conference Proceedings {\bf 140}, 109 (1986).
\bibitem{Jin} C.H. Jin, F. Lin, K. Suenaga, and S. Iijima, Phys. Rev. Lett. {\bf 102}, 195505 (2009).
\bibitem{Morko} H. Morko\c{c}, S. Strite, G.B. Gao, M.E. Lin, B. Sverdlov, and M. Burns, J. Appl. Phys. {\bf 76}, 1363, (1994).
\bibitem{Bruzzone} Bruzzone, S.; Fiori, G. Appl. Phys. Lett. {\bf 99}, 222108 (2011).
\bibitem{Takagi} Takagi, S.-i., Toriumi, A., Iwase, M. Tango, H. IEEE Trans. Electr. Dev. {\bf 41}, 2357 (1994).
\bibitem{Fiori} Fiori, G. Proc. IEEE {\bf 101}, 1653 (2013).
\bibitem{Qiao} Qiao, J.; Kong, X.; Hu, Z.-X.; Yang, F.; Ji, W. Nat. Commun. {\bf 5}, 4475 (2014).
\bibitem{Jia} Jiafeng Xie, Z. Y. Zhang, D. Z. Yang, D. S. Xue, and M. S. Si, J. Phys. Chem. Lett. {\bf 5}, 4073 (2014).
\bibitem{HZ} H. Zeng, C. Zhi, Z. Zhang, X. Wei, X. Wang, W. Guo, Y. Bando and D. Golberg, Nano Lett., {\bf 10}, 5049 (2010).
\bibitem{KS} K. S. Novoselov, D. Jiang, F. Schedin, T. J. Booth, V. V. Khotkevich, S. V. Morozov and A. K. Geim, Proc. Natl. Acad. Sci. U. S. A., {\bf 102}, 10451 (2005).

\end{thebibliography}
\end{document}